\documentclass[12pt, a4paper]{article}

\usepackage{graphics}
 \usepackage{graphicx}
 \usepackage{hyperref}

 \usepackage{epsfig}

\usepackage{amssymb}
\usepackage{amsmath}
 \usepackage{amsthm}

\title{Investigation of coupling geometry and dimerization effects on thermoelectric properties of a $C_{60}$
molecular transistor}

\author{M.˜ Bagheri Tagani, Z.~Golsanamlou, S.~Izadi, and
        H.˜ Rahimpour Soleimani,\\
        \small{Department of physics, University of Guilan, P.O.Box 41335-1914, Rasht, Iran}}

\begin{document}

\maketitle

\begin{abstract}
Thermoelectric properties of a $C_{60}$ molecular transistor are
studied using Green function formalism in linear response regime.
A tight-binding model is used to investigate the effect of the
dimerization and coupling geometry on the electrical conductance,
thermopower, and figure of merit. Increase of the connection
points between the molecule and electrodes results in decrease of
the number of the peaks of the electrical conductance owing to the
interference effects. In addition,  oscillation of the
thermopower is reduced by increase of the connection points. It
is also observed that the kind of carriers participating in the
energy transport is dependent on the coupling geometry. Results
show that the increase of the connection points leads to the
reduction of the figure of merit.
\end{abstract}

\section{Introduction}
\label{Introduction} Molecular junctions and quantum dots (QDs)
have great potential for electronic, spintronic, and energy
conversation applications. Discreteness of energy levels, strong
Coulomb correlations, and interference effects result in the
novel and interesting phenomena such as: negative differential
conductance~\cite{ref1,ref2,ref3,ref4}, ratchet
effect~\cite{ref5}, spin and Coulomb blockade
effects~\cite{ref6,ref7,ref8}, and Kondo
effect~\cite{ref9,ref10,ref11,ref12}. Energy conversion to
electricity is an important challenge which has became a hot
topic in recent years because of the recent advances in
manufacturing nanoscale devices. Strong quantum effects in such
devices result in the violation of the Wiedemann-Franz
law~\cite{ref13,ref14} and as a consequence, increase of the
thermoelectric efficiency. The thermoelectric efficiency is
described by a dimensionless quantity as figure of merit,
$ZT=G_eS^2T/\kappa$, where $G_e$ and $\kappa$ stand for the
electrical and thermal conductances, respectively. $S$ denotes
the thermopower and $T$ is the operating temperature. Unlike
bulky samples, molecular devices can have the figure of merit
higher than unity indicating they can work as good thermoelectric
devices. The thermoelectric experiments can be also used to study
the nature of the transport through molecules. For example, the
positive thermopower shows that the transport is dominant by
holes through the HOMO level, whereas the negative thermopower
indicates the electron participation in the transport through the
LUMO level.

\par The $C_{60}$ molecule is consist of 12 pentagons and 20
hexagons. The energy gap of the molecule is about $2eV$ causing
an insulator-like behavior in the room temperature~\cite{ref15}.
The molecule can be used as a good conductive single molecular
junction because of delocalization of the frontier orbitals of
$C_{60}$. Therefore, a lot of research has been done on the
transport properties of the $C_{60}$ based devices experimentally
and
theoretically~\cite{ref16,ref17,ref18,ref19,ref20,ref21,ref22,ref23,ref24}.
Park et al.~\cite{ref16} fabricated the first individual $C_{60}$
molecular transistor. Their transport experiments revealed the
coupling between the center of mass motion of the molecule and the
electronic degrees of freedom and the step-like behavior of the
current-voltage characteristic. N\'{e}el and
co-workers~\cite{ref17,ref18} studied the influence of the
couplings on the conductance of a $C_{60}$ molecular transistor
and found that the decrease of the distance between the molecule
and the scanning tunneling microscope tip results in the increase
of the conductance. It has been shown that the $C_{60}$ junctions
can have high conductance if they are coupled to the proper
electrodes~\cite{ref23}.

\par Study of the thermoelectric properties of the molecular
junctions has very recently gained a lot of attention from both
experimental and theoretical points of
view~\cite{ref24,ref25,ref26,ref27,ref28,ref29,ref30,ref31,ref32,ref33,ref34,ref35,ref36}.
 Tan and co-workers~\cite{ref28} studied the effect of length and
contact chemistry on the thermoelectric properties of a molecular
junction. They observed the asymmetry of the coupling strength
between the molecule and  electrodes results in the significant
reduction of the electrical conductance, whereas the thermopower
varied by only a few percent. The coupling strength was changed
by  switching the coupling chemistry. In addition, it was found
that the thermopower linearly increases by increment of the
molecule length. Very recently, thermal transport through carbon
nanobuds has been investigated using Molecular dynamics
simulations~\cite{ref36}. Results show that the nanobuds can be
used as good thermal conductor. Balachandran et al.~\cite{ref33}
found that end-group-mediated charge transfer between the
molecule and electrodes plays an important role in the
thermoelectric properties of triphenyl molecules. Their results
also show that the sign of the thermopower is  related to the
HOMO-LUMO energies. Bilan and co-workers~\cite{ref24} studied the
conductance and thermopower of single-molecule junction based on
a $C_{60}$ molecule using the density functional theory within the
nonequilibrium Green function technique. They found that the
junction can be highly conductive and thermopower is negative
because the lowest unoccupied molecular orbital dominates the
charge transport.

\par In this article, the thermoelectric properties of a $C_{60}$
transistor is investigated using Green function formalism in the
linear response regime. The $C_{60}$ can couple to the electrodes
through a single point, a pentagon, or a hexagon. The kind of
coupling can significantly affect the transport properties of the
molecule as it was predicted in ~\cite{ref21}. The kind of the
coupling strongly influences on the transmission shape and with
respect to the the fact that the transmission coefficient is the
most important parameter in the study of the thermopower in the
linear response regime, therefore; the figure of merit of the
$C_{60}$ transistor is strongly dependent on the shape of the
coupling. However, the effect of the coupling geometry and bond
dimerization on the thermoelectric properties of a $C_{60}$ has
not been addressed so far. In the next section, a tight-binding
model is used to describe the $C_{60}$ and electrode Hamiltonian
and the thermoelectric coefficients are presented by means of  the
Green function language.  Sec. III is devoted to the  numerical
results. A brief conclusion is given in Sec. IV .

 \section{Model and Formalism}
We consider a $C_{60}$ molecule coupled to one dimensional
metallic electrodes. The tight-binding approximation with only
one orbital per atom is used to describe the molecule. The
Hamiltonian is given by
\begin{equation}\label{Eq.1}
  H_{mol}=\sum_{i=1}^{60}(\varepsilon_i+eV_{G})d^{\dag}_{i}d_{i}-\sum_{<ij>}t_{ij}d^{\dag}_{i}d_{j},
\end{equation}
where $\varepsilon_{i}$ is the on-site energy of the $i$th
orbital taken as the zero of energy. $V_{G}$ denotes the gate
voltage used to control the energy levels of the molecule.
$d_{i}$ is the annihilation operator destroying an electron in
the orbital $i$. $t_{ij}$ is the hopping matrix element assumed
to be nonzero only between nearest-neighbor orbitals $<ij>$.
Because the $C_{60}$ is composed of the single and double bonds
with different lengths, we set $t_1=2.5$ for single bonds and
$t_2=1.1t_1$ for double bonds~\cite{ref37}.

The electrodes are described as
\begin{equation}\label{Eq.2}
  H_{C}=\sum_{\alpha,i_{\alpha}}\varepsilon_{i_{\alpha}}c^{\dag}_{i_{\alpha}}c_{i_{\alpha}}-\sum_{\alpha
  i_{\alpha}
  j_{\alpha}}t_{i_{\alpha}
  j_{\alpha}}c^{\dag}_{i_{\alpha}}c_{j_{\alpha}},
\end{equation}
where $t_{i_\alpha j_\alpha}=t_1$ is the hopping matrix element of
the electrode $\alpha=L,R$, and
$\varepsilon_{i_{\alpha}}=2t_{i_\alpha j_\alpha}$ is the on-site
energy of the electrode. The coupling between the molecule and
electrodes is given as
\begin{equation}\label{Eq.3}
  H_{T}=\sum_{\alpha i_\alpha j}(t'_\alpha
  c^{\dag}_{i_\alpha}d_{j}+h.c),
\end{equation}
where $t'_\alpha=t_{i_\alpha j_\alpha}/2$ stands for the coupling
strength. We assume that the spin is conserved during the
tunneling from the electrodes to the molecule.

\par The retarded Green function of the molecule coupled to the electrodes
is given as
\begin{equation}\label{Eq.4}
  G^r(\varepsilon,V_G)=[(\varepsilon+i0^{+})I-H_{mol}-\Sigma_{L}(\varepsilon)-\Sigma_R(\varepsilon)]^{-1},
\end{equation}
where the couplings effect is taken into account by the
self-energy functions, $\Sigma_{\alpha}(\varepsilon)$.
$\varepsilon$ denotes the energy of the injected electron from
the leads, and $0^+$ is an infinitesimal value.
$\Sigma_{\alpha}(\varepsilon)$ describes the effect of the
semi-infinite electrodes on the molecule and for one dimensional
electrodes is given as~\cite{ref38}
\begin{equation}\label{Eq.5}
  \Sigma_{\alpha}(\varepsilon)=-t'^2_{\alpha}g_{\alpha}(\varepsilon)=-\frac{t'^2_{\alpha}}{t}e^{ika},
\end{equation}
where $t=t_{i_\alpha j_\alpha}$,
$ka=acos(1-(\varepsilon-\varepsilon_\alpha)/2t)$, and
$\varepsilon_\alpha=\varepsilon_{i_{\alpha}}$. The real part of
the self-energy shifts the position of the energy levels of the
molecule, whereas its imaginary part results in the  broadening of
the density of  states of the molecule and the finite lifetime of
the electron in the molecule. The coupling between the $C_{60}$
and electrodes depends on the orientation of the molecule so that
one carbon atom, a pentagon or a hexagon may couple to the
electrodes. The coupling geometry significantly affects the
self-energy matrix so that it has only one nonzero element for
single atom connection, 25 nonzero elements for a pentagon and 36
nonzero elements for a hexagon coupling. The more details about
the coupling geometry can be found in ~\cite{ref21}.

\begin{figure}
\begin{center}
\includegraphics[height=150mm,width=100mm,angle=0]{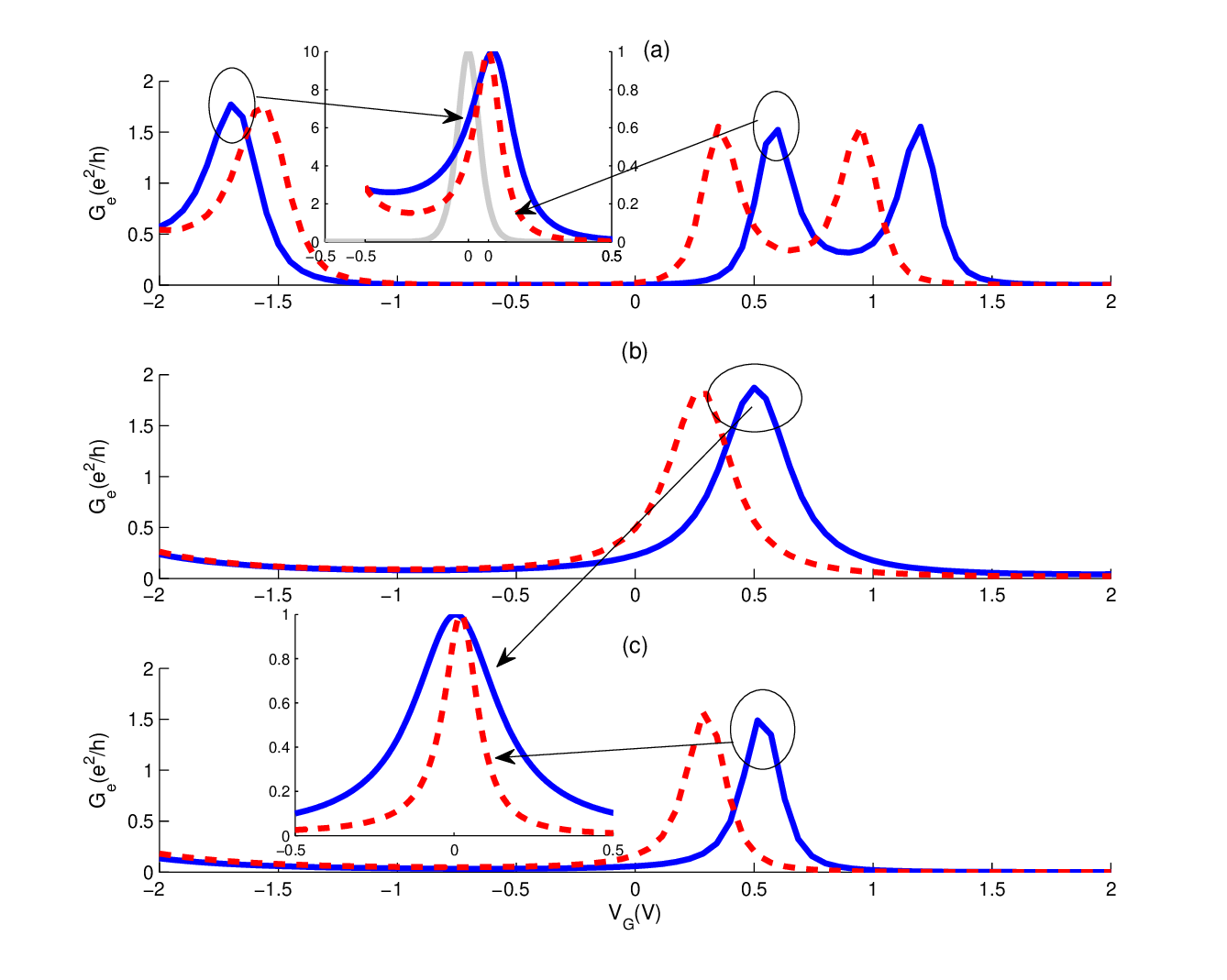}\nonumber
\caption{Electrical conductance  (a) single point coupling, (b)
five points coupling, and (c) six points coupling for $t_2=t_1$
(solid line) and $t_2=1.1t_1$ (dashed line). Upper panel shows
the transmission coefficient at $V_G=-1.7$  (solid) and $V_G=1.5$
(dashed). The fermi derivative is plotted in gray. Lower panel
shows the transmission coefficient of the five atom (solid) and
six atom (dashed) connections for voltage gates indicated by
circle.
}\label{fig:1}       
\end{center}
\end{figure}

\par In order to compute the charge and heat currents, the Keldysh
nonequilibrium Green function formalism is used. The charge and
heat currents are given as follows~\cite{ref38,ref39}
\begin{subequations}\label{Eq.6}
\begin{align}
I&=\frac{2e}{h}\int d\varepsilon
[f_{L}(\varepsilon)-f_{R}(\varepsilon)]T(\varepsilon),\\
Q&=\frac{2}{h}\int d\varepsilon (\varepsilon-\mu_{L})
[f_{L}(\varepsilon)-f_{R}(\varepsilon)]T(\varepsilon),
\end{align}
\end{subequations}
where
$f_{\alpha}(\varepsilon)=[1+exp((\varepsilon-\mu_\alpha)/kT_\alpha)]^{-1}$
is the Fermi distribution function of the electrode $\alpha$,
$\mu_\alpha$ and $T_\alpha$ denote, respectively, the chemical
potential and temperature of the electrode $\alpha$.
$T(\varepsilon)=Tr[\Gamma_L(\varepsilon)G^{r}(\varepsilon)\Gamma_R(\varepsilon)G^{a}(\varepsilon)]$
is the transmission coefficient and
$\Gamma_\alpha=-2Im(\Sigma_\alpha)$ is the coupling matrix.

\par We investigate the thermoelectric properties of the $C_{60}$
molecule in the linear response regime in which the charge and
heat currents are expressed in terms of the applies temperature
difference, $\Delta T$, and induced voltage drop, $\Delta V$, to
first order according to
\begin{subequations}\label{Eq.7}
\begin{align}
I&=e^2L_0\Delta V+\frac{e}{T}L_1 \Delta T,\\
Q&=eL_1\Delta V+\frac{1}{T}L_2 \Delta T,
\end{align}
\end{subequations}
where $L_n$ are integrals of the form
\begin{equation}\label{Eq.8}
  L_n=-\frac{1}{h}\int d\varepsilon (\varepsilon-\mu)^n \frac{\partial
  f}{\partial \varepsilon} T(\varepsilon),
\end{equation}
where $\mu$ is the chemical potential of the leads in the
equilibrium. Thermopower is the ratio of the voltage drop to the
applied temperature difference under condition that the current
vanishes $I=0$, therefore, $S=-\frac{1}{eT}\frac{L1}{L0}$. The
electrical, $G_e$, and thermal, $\kappa$, conductances are given
as
\begin{subequations}\label{Eq.9}
\begin{align}
  G_e&=e^2L_0,\\
  \kappa&=\frac{1}{T}[L_2-\frac{L_1^2}{L_0}].
\end{align}
\end{subequations}
In the following, we analyze the dependence of the electrical and
thermal conductances, thermopower and the figure of merit on the
dimerization, coupling geometry and temperature. To compute the
figure of merit, we neglect the lattice thermal conductance which
is usually small in nanostructures.

\section{Results and Discussion}

\begin{figure}
\begin{center}
\includegraphics[height=150mm,width=110mm,angle=0]{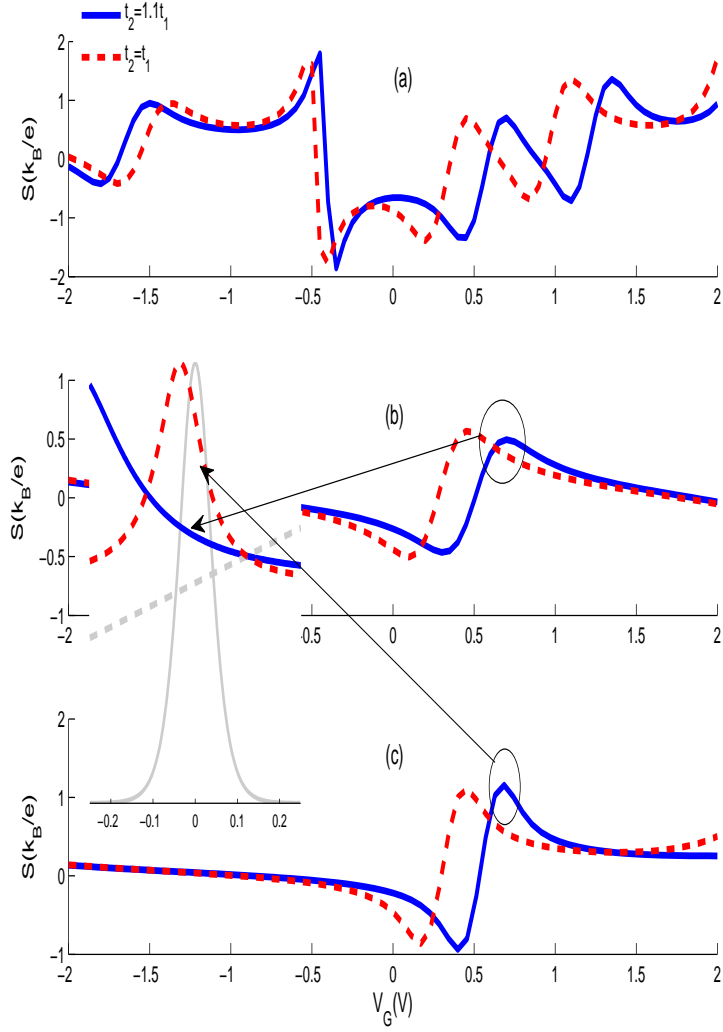}\nonumber
\caption{Thermopower versus gate voltage for (a) single atom
connection, (b) five atom connections, and (c) six point
connections at $t_2=t_1$ (solid) and $t_2=1.1t_1$ (dashed). Inset
shows the transmission function of the five point (solid) and six
point (dashed) connections. The Fermi derivative and
$\varepsilon-\mu$ are shown by gray solid and dashed
lines, respectively. $T=300K$. }\label{fig:2}       
\end{center}
\end{figure}

Figure 1 shows the electrical conductance as a function of gate
voltage for different coupling geometries. In single atom contact,
the conductance peaks are exactly located at the molecular energy
levels as it was expected. There are three peaks denoting  HOMO,
LUMO, and LUMO+1, respectively. The gap is about $2eV$ which is
consistent with previous results. The dimerization affects HOMO
and LUMO peaks in different manners   so that the peak located at
HOMO is shifted toward right, whereas the one located at LUMO is
moved to the left when $t_1=t_2$. The eigenvalues of Hamiltonian
of the $C_{60}$ are strongly dependent on $t_1$, and $t_2$,
therefore, such change is predictable. Our analysis shows that
the gap is reduced up to $0.35 eV$ when $t_1=t_2$. In addition,
it is observed the peak located at HOMO is slightly higher than
the ones located at LUMO and LUMO+1. With respect to the fact the
electrical conductance is directly related to the transmission
coefficient, the difference arises from the changes of the
transmission. In the upper panel, we plot the transmission
coefficient at $V_G=-1.7$, HOMO, and $V_G=1.5$, LUMO. As it is
observed the transmission is wider in the HOMO and the difference
in the width of the transmission coefficient gives rise to the
difference in the height of $G_e$.
\par
Unlike single atom contact, the electrical conductance exhibits
just one peak located at LUMO in the case of connection to five or
six carbon atoms. The decrease of the number of the conductance
peaks is a direct result of interference effects. In fact, with
increase of the crossing channels for injection of electrons from
electrodes to the molecule, the electron waves may suffer  a
destructive or constructive interference. The destructive
interference leads to the vanishing of the transmission peaks and
as a result, the electrical conductance peaks are disappeared.
Moreover, it is observed that the position of the peak in five or
six atom connections is slightly different from the single atom
connection. The difference comes from the real part of the self
energy. Real part of self energy shifts the molecular energy
levels and such shift becomes more pronounced with increase of
connection points. As single point case, the peak is moved toward
left when $t_1=t_2$, however, the height of peak is constant. In
addition, it is observed that the electrical conductance has
higher peak in five point connections because the transmission
coefficient is wider in five point connections, see lower panel.
\begin{figure}
\includegraphics[height=130mm,width=110mm,angle=0]{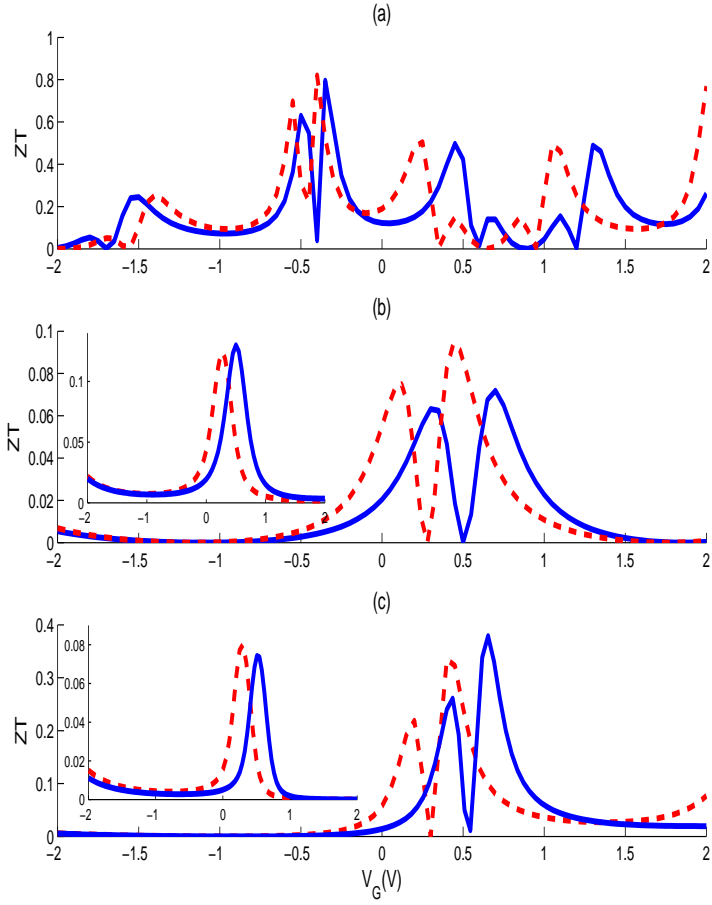}\nonumber
\caption{ $ZT$ versus gate voltage in (a) single atom, (b) five
atom and (c) six atom connections for $t_2=t_1$ (solid) and
$t_2=1.1t_1$ (dashed). Upper panel shows the thermal conductance
of the
five point connections case, while the lower panel shows the thermal conductance of the six atom connections.}\label{fig:3}       
\end{figure}

\par The dependence of thermopower on gate voltage  is plotted in Fig.
2. The change of the electron population of the molecule results
in the oscillation of the thermopower which has been extensively
reported for the quantum dots and
molecules~\cite{ref40,ref41,ref42}. The sign of the thermopower
determines the kind of the carrier responsible for the transfer
of the current and energy. In the single point contact, the
thermopower has five zeros which three of them are located at
resonance energies and others are in the electron-hole symmetry
points. The thermopower varies sharper in the symmetry points. In
the resonance energies the thermopower is zero because the
temperature difference cannot induce a net current. In fact,
electrons can tunnel from the hotter and colder electrodes to the
molecule without needing to the thermal energy. In symmetry
points, electrons and holes participate in the transfer of charge
and energy with the same weight. Electrons and holes carry the
charge in the opposite directions so the net current is zero. In
this points, the thermal conductance is maximum because they
carry the energy in the same direction. The sign of the
thermopower changes in the vicinity of the resonance and symmetry
energies. For gate voltages lesser than resonance, the
thermopower is positive (note that the thermopower in the unit of
$k_B/e$ is negative due to negative electron charge.) indicating
the electrons of the hotter lead carry the current, whereas for
gate voltages lesser than symmetry energies, the thermopower is
negative because the holes carry the current. Such behavior was
previously reported for systems composed of single and double
QDs~\cite{ref41,ref43}. It is interesting to note that the
thermopower is asymmetry because of the electron-hole symmetries.
The dimerization changes the position of the resonance and
symmetry points. In addition, the magnitude of the thermopower
slightly changed.
\par The increase of the connection points significantly reduces
the oscillation of the thermopower. It comes from the fact that
the lesser molecular energy levels are involved in the
thermoelectric transport due to the destructive interference
effects. Results show that the magnitude of the thermopower is
more in the six point contacts than the five point contacts. The
difference is a direct result of the change of the transmission
in the vicinity of the chemical potential of the electrodes. As
it is shown in the inset, the transmission coefficient of the six
point contact is more in the range of
$\partial{f}/\partial{\varepsilon}$, therefore, $L_1$ increases
and as a consequence, the thermopower is increased.

\par Figure 3 shows the figure of merit as a function of gate
voltage for different coupling geometries. In single point
connection, $ZT$ has a lot of peaks with significant magnitudes
in a wide range of the gate voltage. Increase of the number of
connection points results in the decrease of the number of peaks.
This reduction results from the destructive interference
previously discussed. The change of the strength of the bonding
does not affect the magnitude of the $ZT$ and just changes the
position of the peaks. In the five and six point connections, the
dimerization influences on the magnitude of the $ZT$. In five
point connections, the figure of merit increases when $t_2>t_1$
owing to the increase of the thermopower and the decrease of the
thermal conductance. The thermal conductance of the five point
connections is plotted in the upper panel. It is so interesting
to note that the $ZT$ of the six point connections is increased
when $t_1=t_2$. In this case, although the thermopower is
slightly decreased, the electrical  and thermal conductances are
increased with different rates so that the figure of merit is
enhanced. The behavior of the thermal conductance of the six point
connections is plotted in the lower panel.
\begin{figure}
\includegraphics[height=140mm,width=100mm,angle=0]{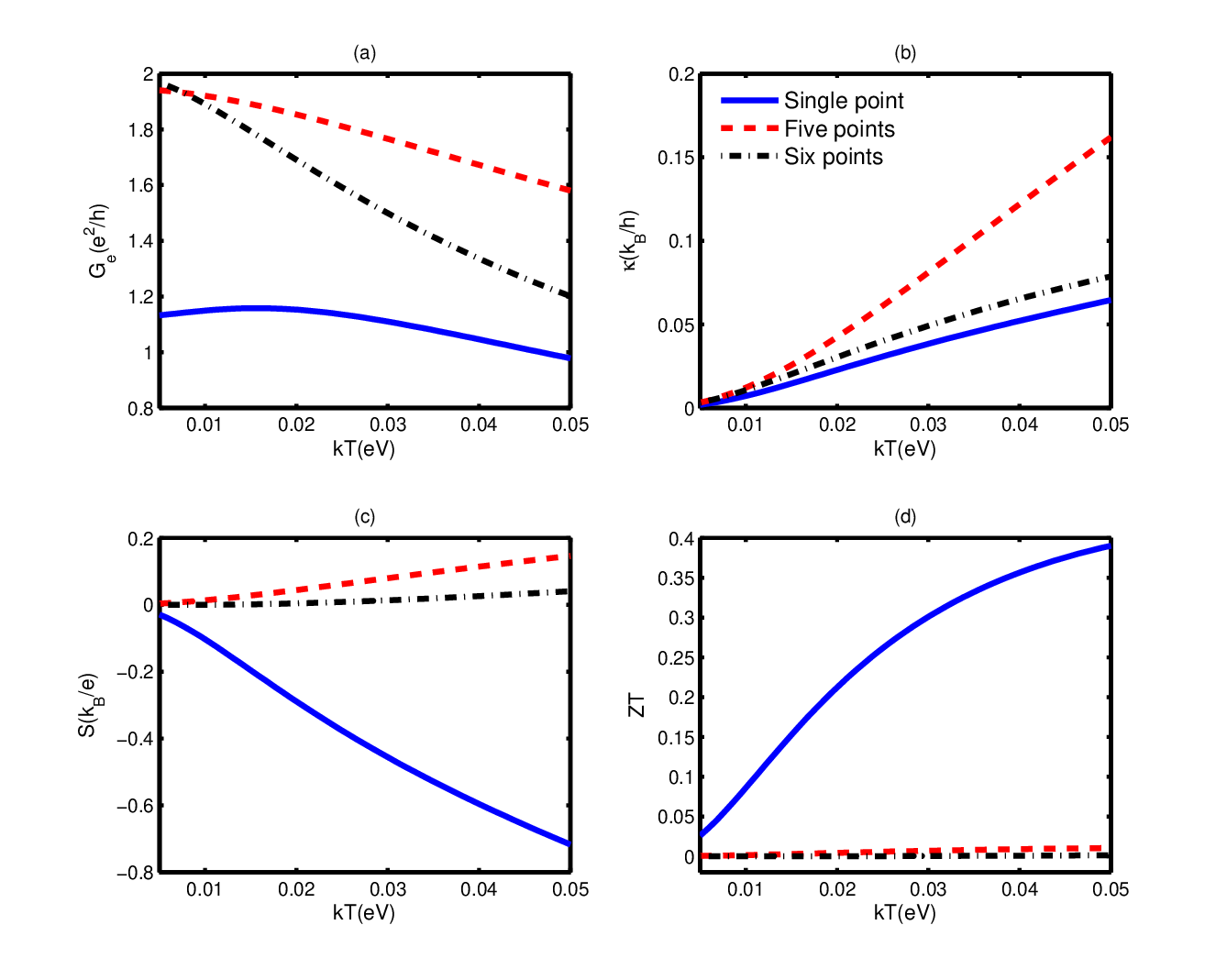}\nonumber
\caption{(a) Electrical conductance, (b) thermal conductance, (c) thermopower, and (d) figure of merit against temperature.}\label{fig:4}       
\end{figure}
\par Dependence of the thermoelectric coefficients on the
temperature is plotted in Fig. 4. We set $V_G=0.3 V$. As it is
observed the coupling geometry strongly affects the coefficients.
The increase of temperature results in the broadening of the Fermi
derivative and increase of the electron population in the
molecule, nevertheless, the transmission coefficient is
independent of temperature. It is worth noting that the Green
function becomes a function of the electron density if the
Coulomb correlations are taken into account. Fig. 5a and b shows
the transmission coefficients for different coupling geometries
and the Fermi derivative for different temperatures, respectively.
\begin{figure}
\includegraphics[height=100mm,width=100mm,angle=0]{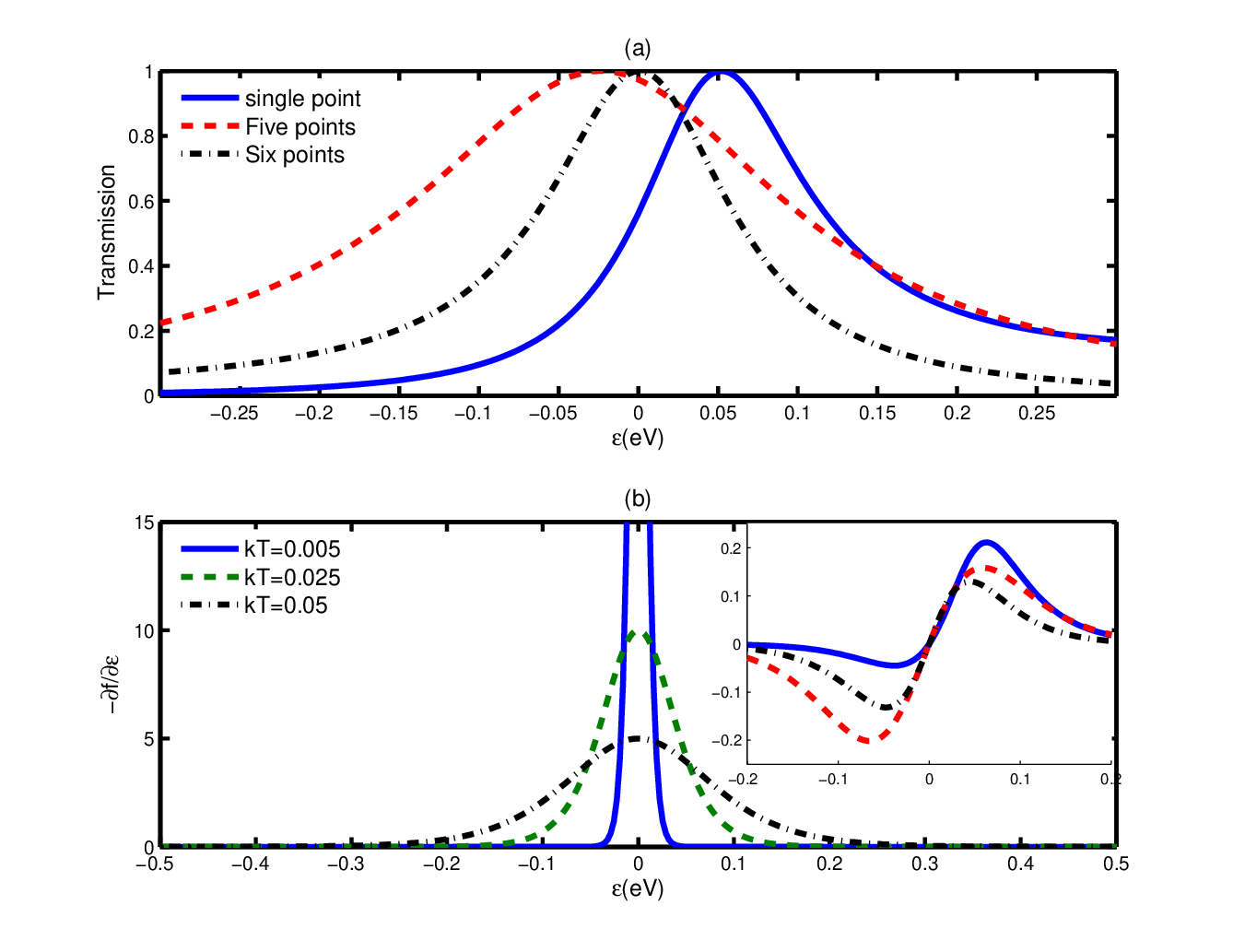}\nonumber
\caption{ (a) Transmission coefficient and (b) Fermi derivative
versus energy. Inset shows
$T(\varepsilon)(\varepsilon-\mu)f'(\varepsilon)$}\label{fig:5}       
\end{figure}
The transmission coefficient is a Lorentz-like function in the
vicinity of the chemical potential of the electrodes, see Fig.
5a, whose center and width are related to the coupling geometry.
The center of the transmission coefficient of the six point
connections is exactly located at the Fermi energy, therefore,
its electrical conductance is more than others in low
temperatures. With increase of temperature, the electrical
conductance of the five point connections is more than others
because its transmission coefficient is wider. The electrical
conductance of the five and six point connections is monotonically
decreased by increase of temperature because of the reduction of
the height of the Fermi derivative. The  dependence of the
electrical conductance of single point on the temperature is
non-uniform so that first, it increases and then decreases. The
initial increase comes from the fact that the center of the
transmission coefficient is slightly far from the Fermi energy,
therefore, the increase of temperature brings more parts of the
transmission in the nonzero energy of the Fermi derivative. The
increase of temperature significantly reduces the height of the
Fermi derivative, see Fig. 5b, so the electrical conductance
decreases in high temperatures. Results show that the thermal
conductance increases with increase of the temperature because
the carriers convey more thermal energy. The increase is more
remarkable in the five point connections resulting more
broadening of the transmission in this case.
\par Fig. 4c shows the thermopower versus temperature. It is so
interesting that the kind of carriers participating in the
thermoelectric transport is different for single and multi
couplings. In single point connection, electrons participate in
the transport of the current and heat flux while the holes play
the main role in multi couplings. It occurs due to the shape and
more specially, the position of the center of the transmission
peak. The center of peak is in positive energies for single
point, therefore, electrons thermally excited and above the
chemical potential of the leads can tunnel through the molecule.
In five and six point connections, the center is in negative
energies, therefore, more holes below the Fermi energy of the
leads can enter the molecule. In addition, the magnitude of the
thermopower in single point is much more other cases owing to the
more energy distance of the center of the transmission from the
Fermi energy. For more clarity, we plot
$T(\varepsilon)(\varepsilon-\mu)f'(\varepsilon)$ in the Inset of
the Fig. 5. One can see that this quantity is asymmetric around
the Fermi energy so that, its positive part is more in single
point and its negative part is more in other cases. The figure of
merit is plotted in Fig. 4d. The  thermopower is the more
important quantity for the $ZT$, therefore, although the
electrical conductance reduces in the single point connection,
the figure of merit significantly increases  because of the
magnitude of the thermopower. Increase of temperature results in
the increase of the figure of merit because the magnitude of the
thermopower is enhanced with increase of $kT$.

\section{Summary}
In this article, we have analyzed the thermoelectric properties
of a $C_{60}$ molecular transistor in the linear response regime.
A tight-binding model within the Green function formalism is used
to compute the electrical conductance, thermopower, and figure of
merit. Results show that the coupling geometry strongly control
the thermoelectric properties of the device. Increase of the
number of coupling points results in the decrease of the
oscillation of the thermopower, and decrease of the figure of
merit. It is a direct result of the interference effects. In
addition, the kind of the carriers participating in the
thermoelectric transport is dependent on the coupling geometry
and temperature.

\end{document}